\documentclass[conference]{IEEEtran}
\IEEEoverridecommandlockouts
\usepackage{cite}
\usepackage{amsmath,amssymb,amsfonts}
\usepackage{algorithmic}
\usepackage{graphicx}
\usepackage{textcomp}
\usepackage{xcolor}

\usepackage{algorithm}
\usepackage{hyperref}
\usepackage{multirow}

\def\BibTeX{{\rm B\kern-.05em{\sc i\kern-.025em b}\kern-.08em
    T\kern-.1667em\lower.7ex\hbox{E}\kern-.125emX}}

\renewcommand{\footnoterule}{%
  \kern -3pt
  \hrule width 0.2\textwidth
  \kern 2.6pt
}

\begin{document}

\title{Dynamic Multi-Species Bird Soundscape Generation with Acoustic Patterning and 3D Spatialization}


\author{
\IEEEauthorblockN{Ellie L. Zhang}
\IEEEauthorblockA{
 \textit{IntelliSky}\\
McLean, USA \\
elzhang@intellisky.org}
 \and 
\IEEEauthorblockN{Duoduo Liao}
\IEEEauthorblockA{
 \textit{George Mason University}\\
Fairfax, USA \\
dliao2@gmu.edu}
 \and
\IEEEauthorblockN{Callie C. Liao}
\IEEEauthorblockA{
\textit{Stanford University}\\
Stanford, USA \\
ccliao@stanford.edu}
}

\maketitle

\begin{abstract}

Generation of dynamic, scalable multi-species bird soundscapes remains a significant challenge in computer music and algorithmic sound design. Birdsongs involve rapid frequency-modulated chirps, complex amplitude envelopes, distinctive acoustic patterns, overlapping calls, and dynamic inter-bird interactions, all of which require precise temporal and spatial control in 3D environments. Existing approaches, whether Digital Signal Processing (DSP)-based or data-driven, typically focus only on single species modeling, static call structures, or synthesis directly from recordings, and often suffer from noise, limited flexibility, or large data needs. To address these challenges, we present a novel, fully algorithm-driven framework that generates dynamic multi-species bird soundscapes using DSP-based chirp generation and 3D spatialization, without relying on recordings or training data. Our approach simulates multiple independently-moving birds per species along different moving 3D trajectories, supporting controllable chirp sequences, overlapping choruses, and realistic 3D motion in scalable soundscapes while preserving species-specific acoustic patterns. A visualization interface provides bird trajectories, spectrograms, activity timelines, and sound waves for analytical and creative purposes. Both visual and audio evaluations demonstrate the  ability of the system to generate dense, immersive, and ecologically inspired soundscapes, highlighting its potential for computer music, interactive virtual environments, and  computational bioacoustics research.

\end{abstract}


\section{Introduction}

Generating dynamic bird vocalizations and scalable multi-species soundscapes is a long-standing challenge in computer music and algorithmic sound design as well as bioacoustics research. Particularly, birdsongs exhibit rapid, highly structured frequency-modulated chirps, intricate amplitude envelopes, distinctive acoustic patterns, overlapping calls, rich multi-voice interactions, and dynamic inter-bird behaviors, all of which contribute to complex and continuously evolving acoustic textures~\cite{farina2017ecoacoustics, Zysman2005, CatchpoleSlater2008}. Accurately reproducing these features in three-dimensional (3D) space requires precise control over oscillators, modulation schemes, temporal envelopes, and the simulation of spatial and dynamic interactions among multiple vocal sources~\cite{Mindlin2017}. 

Although prior work has explored the synthesis of bird calls using Digital Signal Processing (DSP)~\cite{DAVIS1986171} or biophysically realistic models of the vocal apparatus\cite{Coen2007, PhysRevE.81.031927}, these approaches typically focus on 1) a single species, 2) static or hand-designed call structures, or 3) synthesis directly from real recordings \cite{Zysman2005}, which are difficult to scale to multi-species interaction environments. Consequently, fully algorithmic, dynamic, scalable multi-species soundscape generation remains a challenge due to overlapping calls, varying movement, and the need for realistic 3D perception \cite{Mindlin2017}.

Recent advances in computational bioacoustics \cite{ Bonada2016, Stowell2022ComputationalBioacoustics} and immersive technologies \cite{ Chakravarty2020Directional, Tong2025MultimodalAcousticFields}, such as Virtual Reality (VR)~\cite{XRAudio2025, spatialAdioVRReview2023}, have enabled detailed sound synthesis and environmental audio modeling. However, most existing methods still neglect species-specific acoustic variability and scalability. Some rely heavily on static recordings that cannot be flexibly adapted to new contexts. These limit their usefulness for interactive applications, scalable virtual environments, or generated sound design. 

Moreover, recent approaches predominantly rely on recorded bird vocalizations or data-driven methods \cite{Stowell2022ComputationalBioacoustics, Lecomte2023}, including machine learning models, which present significant limitations. Collecting field recordings is labor-intensive, often contaminated by background noise, and provides limited flexibility for generating novel or controllable soundscapes. Data-driven models, on the other hand, require large and diverse training datasets and are inherently constrained by the scope and variability of the recorded material. Furthermore, biological audio datasets are often small or imbalanced, as species-specific or rare calls are difficult to obtain.

To address these challenges, we propose a novel, fully algorithmic and DSP-based framework to generate dynamic, scalable multi-species bird soundscapes that integrates species-specific temporal and spectral patterns, chirp-level dynamics, overlapping calls, inter-bird variability, and 3D spatialization. This framework enables controllable, scalable, and dynamic simulation of multi-species bird activity in 3D space, capturing both acoustic and behavioral diversity without relying on recordings or training data. Moreover, the framework includes visualization tools for bird trajectories, spectrograms, activity timelines, and waveforms, enabling quantitative analysis of spatial and temporal patterns in the simulated soundscape. To our knowledge, no prior work combines fully algorithmic generation with species-specific acoustic patterning, multi-species interaction, and realistic 3D spatialization. This approach expands opportunities for interactive audio, computer music composition, immersive virtual environments, and bioacoustic research. 

The main contributions of this work are:
\begin{itemize}
    \item A fully algorithmic DSP-based framework for dynamic multi-species 3D soundscape generation, capturing species-specific frequency and temporal chirp features without relying on recordings or machine learning, thereby expanding controllable sound design in computer music and computational bioacoustics.
    \item A real-time 3D spatialization and movement model that integrates dynamic trajectories, vectorized panning, and distance-based attenuation, enabling spatial motion to function as a musically expressive parameter in immersive audio environments.
    \item A multi-species interaction system that supports overlapping calls, temporal interplay, and 3D spatialization, creating a novel algorithmic approach for generating complex, dynamic, and scalable bird soundscapes.
    \item A comprehensive set of mathematical formulations provided for 3D bird-soundscape generation, unifying chirp-level signal synthesis, 3D acoustic spatialization, and multi-species soundscape modeling in a fully reproducible framework. 
    \item Visualization of trajectories, spectrograms, chirp activity timelines, and comparative analysis that support analytical evaluation and creative exploration of algorithmic sound processes and pedagogical use.
    \item An extensible platform that bridges computer music, algorithmic composition, virtual environments, and ecological audio simulation, providing artists and researchers  with a flexible system for generating controlled and ecologically-informed 3D soundscapes.
    
\end{itemize}

\section{Related Work}

\subsection{Foundations of Bird Vocalization}
Early studies have provided a biological and ecological basis for understanding bird vocalizations. Bradbury and Vehrencamp \cite{Bradbury1998Principles} formalized the principles of song structure, while Sueur et al. \cite{Sueur2008} examined soundscape ecology using passive acoustic monitoring. These foundational studies inform computational modeling approaches that aim to capture the temporal and spectral characteristics of bird calls. Traditional parametric and signal-based modeling techniques include chirp generation, linear and exponential frequency sweeps, and amplitude envelope shaping \cite{DAVIS1986171}. Such methods provide interpretable models for individual calls but are often limited in capturing complex interactions in multi-species environments. More recently, although deep learning approaches have been largely applied for species classification and call detection \cite{Ghani2023BirdsongEmbeddings, Ghani2025BirdsongTransfer}, generative modeling for realistic simulation remains relatively underexplored. In contrast, our approach simulates dynamic and scalable soundscapes for multiple bird species in 3D environments with pure algorithms, incorporating species-specific acoustic patterns, overlapping calls, inter-species variability, and 3D spatial trajectories.

\subsection{Birdsong Synthesis Techniques}
Birdsong synthesis has been approached from multiple perspectives in computer music. Single-voice or monophonic methods, often based on physical modeling, include Frequency Modulation (FM) synthesis \cite{Roads1996}, additive synthesis \cite{Loy2006}, and nonlinear modulation techniques \cite{Farnell2010, Uncini2022}. These approaches can produce expressive bird-like timbres but generally focus on isolated sounds and do not account for dynamic, scalable multi-species soundscapes or spatial movement. Recently, data-driven and sample-based methods leveraging recorded bird calls or machine learning models have been employed to generate more naturalistic sequences~\cite{song2025bioacousticGen, nime2025_12}. Although these methods can reproduce realistic sounds, they typically require large datasets, suffer from noise contamination, and offer limited control over fine-grained temporal and spatial variation \cite{Sueur2008, NAPIER2024124220}. Consequently, current approaches struggle to capture overlapping calls, inter-bird interactions, or fully 3D immersive soundscape. By contrast, our method generates clear, noise-free birdsongs with multiple independently moving birds per species, fully controllable chirp patterns, and 3D spatialization, all achieved through scalable, algorithmic DSP without relying on prior recordings or large training datasets.

\subsection{3D Spatialized Bird Soundscapes}

For 3D audio rendering~\cite{XRAudio2025, spatialAdioVRReview2023}, spatialization techniques including amplitude panning, distance attenuation, and trajectory-based movement are widely used in gaming and VR \cite{Begault1994, Zhao_2019_ICCV,  Li2018360Audio, Taylor2009RESound}. However, few studies have applied these methods to ecological simulations~\cite{xu2020ecological}. Previous digital soundscape simulations have typically relied on recorded calls or simple tone synthesis \cite{SoundscapeEcology2011}. Moreover, very little published work combines fully algorithmic bird soundscape generation with 3D spatial sound rendering. In particular, integrating species-specific acoustic models into dynamic, multi-species 3D soundscapes remains an open challenge. In this context, our work innovatively combines digital sound techniques with species-specific acoustic patterns, multi-bird movement, and interspecies interactions to simulate dynamic 3D birdsong soundscapes, offering a promising direction for both ecological and musical applications.

\section{Methodology}

\begin{figure*}[t]
\centering
\includegraphics[width=0.9\textwidth]{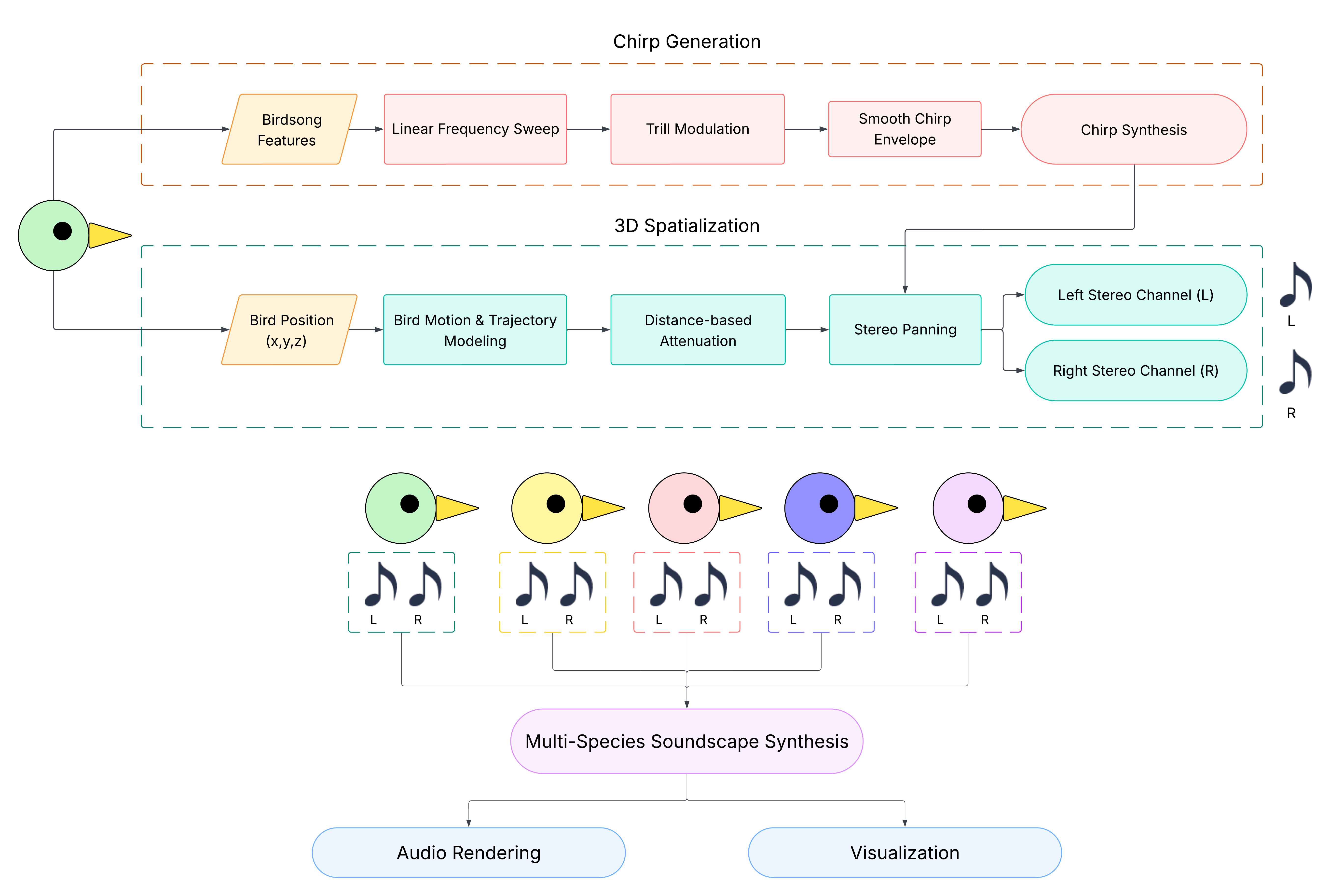}
\caption{The framework of multi-species 3D bird soundscape generation.}
\label{fig:birdsong_framework}
\end{figure*}

We present an efficient framework for synthesizing dynamic, multi-species bird soundscapes with realistic acoustic and spatial properties. The approach consists of four main stages: (1) \textit{chirp generation}, using DSP-based models to simulate species-specific vocalizations with variable temporal and spectral features; (2) \textit{3D spatialization}, assigning dynamic trajectories to produce realistic movement and immersive spatial perception; (3) \textit{multi-species bird soundscape synthesis}, combining individual calls to simulate multi-species multi-bird interactions and musically engaging textures; and (4) \textit{audio rendering and visualization evaluation}, including bird trajectories, spectrograms, activity timelines, comparative analysis of 3D audio waveforms, and WAV audio files for compositional or analytical use. This integrated DSP pipeline enables ecologically realistic and musically expressive auditory simulations. 

Figure \ref{fig:birdsong_framework} outlines the framework, and the following sections detail each stage, including the mathematical formulations underlying call synthesis, feature modeling, and spatialization.

\subsection{Chirp Generation}

Chirp signals are synthesized through DSP \cite{oppenheim2010discrete}, employing a linear frequency sweep combined with a windowed amplitude envelope and a low-amplitude sinusoidal modulation (trill). The frequency transitions from a start ($f_0$) to an end ($f_1$) frequency over a fixed duration, while the envelope ensures smooth onset and offset, and the trill introduces rapid frequency fluctuations, enabling precise control on the chirp's temporal and spectral properties. Each species is modeled with distinct chirp characteristics, including frequency ranges, trill rate, duration, repetition rate, and pause.

\subsubsection{Linear Frequency Sweep}

The linear frequency sweep  \(f(t)\)  defines the pitch of a bird chirp (i.e., the perceived frequency), specifying how it evolves linearly evolves over time, from a start pitch to an end pitch over a duration:
\begin{equation}
    f(t) = f_0 + (f_1 - f_0)\frac{t}{T}, \quad 0 \leq t \leq T
\end{equation}
where $f_0$ and $f_1$ are the start and end frequencies, and $T$ is the chirp duration. 

\subsubsection{Trill Modulation}
To synthesize realistic bird-like chirps, the instantaneous frequency \(f_{\text{trill}}(t)\) combines a smooth linearly varying pitch component (the sweep) with rapid oscillations (the trill). Integrating this instantaneous frequency over time produces the corresponding phase of the sine wave, thereby generating a frequency-modulated waveform. The function \(f_{\text{trill}}(t)\)  is defined as:

\begin{equation}
    f_{\text{trill}}(t) = f(t) \cdot (1 + a \cdot  \sin(2 \pi r t))
\end{equation}
where $r$ is the trill rate and $a$ is the modulation amplitude (dimensionless, typically \( 0 < a < 1 \)).

\subsubsection{Smooth Chirp Envelope}

The chirp envelope, controlling amplitude over time, is shaped using a smooth window function to prevent clicks and emulate the gradual amplitude modulation of natural bird vocalizations. Mathematically, it is defined as:
\begin{equation}
    A(t) = \sin^n\left(\pi \frac{t}{T}\right)
\end{equation}
where $n$ controls fade smoothness. This envelope smoothly increases the amplitude from zero at $t = 0$ to a maximum at $t = T/2$ and decreases back to zero at $t = T$, ensuring a natural fade-in and fade-out of the chirp.

\subsubsection{Chirp Synthesis}

A chirp is a frequency-modulated signal whose instantaneous frequency varies over time. To generate a realistic  chirp, we combine a linear frequency sweep, a trill modulation, and a smooth amplitude envelope. Mathematically, the chirp signal for a certain species is defined as:

\[
s(t) = A(t) \cdot \sin\left( 2 \pi \int_0^t f_{\text{trill}}(\tau) \, d\tau \right)
\]
where \( f_{\text{trill}}(t) \) is the instantaneous frequency and \( A(t) \) is the amplitude envelope. The integral term defines the instantaneous phase by accumulating frequency variations over time, producing the continuous frequency variations typical of bird calls. The amplitude envelope \( A(t) \) is applied to control the onset and offset of the signal, reducing transient artifacts and approximating the smooth intensity variations characteristic of natural bird vocalizations. This formulation enables the generation of naturalistic chirps with controlled frequency sweeps and amplitude modulation.

\subsection{3D Spatialization}

To realistically simulate birds within a 3D auditory scene, stereo signals are generated by applying distance-based attenuation and azimuth-dependent panning. Bird positions are dynamically updated over time, and these spatial transformations are computed for each chirp segment to maintain continuous and perceptually accurate spatialization.

\subsubsection{Bird Motion and Trajectory Modeling}

Bird positions are updated per chirp using sinusoidal trajectories to emulate natural motion:

\begin{equation}
\begin{aligned}
x(t) &= x_0 + \Delta_x \sin(2\pi f_x t), \\
y(t) &= y_0 + \Delta_y \sin(2\pi f_y t), \\
z(t) &= z_0 + \Delta_z \sin(2\pi f_z t),
\end{aligned}
\end{equation}

where \(\Delta_{x,y,z}\) control movement amplitude and \(f_{x,y,z}\) are low-frequency modulations along each axis. This approach allows larger, smooth movements and generates a more dynamic and realistic soundscape.

\subsubsection{Distance-based Attenuation}

The listener is assumed to be at the origin. The distance \(d(t)\) and azimuth angle \(\theta(t)\) in the horizontal (x-y) plane of each bird relative to the listener are computed as:

\begin{equation}
d(t) = \sqrt{x(t)^2 + y(t)^2 + z(t)^2},
\end{equation}
\begin{equation}
\theta(t) = \arctan2\big(y(t), x(t)\big).
\end{equation}

The distance \(d(t)\) is used to compute the amplitude attenuation $\alpha(t)$ according to the inverse distance law:

\begin{equation}
\alpha(t) = \frac{1}{1 + d(t)},
\end{equation}

\subsubsection{Stereo Panning}

The azimuth angle \(\theta(t)\) determines the stereo panning using the equal-power stereo formula:

\begin{equation}
\begin{aligned}
L(t) &= s(t) \cdot \alpha(t) \cdot \cos\!\Big(\frac{\theta(t)}{2}\Big), \\
R(t) &= s(t) \cdot \alpha(t) \cdot \sin\!\Big(\frac{\theta(t)}{2}\Big),
\end{aligned}
\end{equation}

where \(s(t)\) denotes the chirp signal and \(L(t)\) and \(R(t)\) represent the left and right stereo channels, respectively. This approach achieves perceptually accurate spatialization by combining distance-based attenuation with azimuth-dependent panning, enabling listeners to perceive both lateral and depth cues of each moving bird within the synthesized soundscape \cite{7911385}.

\subsection{Multi-Species Bird Soundscape Synthesis}

The stereo signals from all birds are summed over time. To prevent clipping, the resulting waveform is normalized. Each species and individual bird emits multiple chirps according to its rate parameters, producing an evolving multi-species soundscape.

For \( N \) birds of \( S \) species, each track \( x_{s,n}(t) \) is generated and summed:
\begin{equation}
S(t) = \sum_{s=1}^{S} \sum_{n=1}^{N_s} x_{s,n}(t)
\end{equation}
Normalization ensures that the combined signal stays within [-1,1]. Chirp times are recorded to plot activity timelines.

The final soundscape is computed by summing individual bird tracks, with normalization to prevent clipping:
\begin{equation}
    S_{normalized}(t) = \frac{S(t)}{\max|S(t)|}
\end{equation}

\begin{table*}[t]
\caption{Bird Species Parameters}
\centering
\begin{tabular}{|c|c|c|c|c|}
\hline
Species & Freq Range (Hz) & Chirp Duration Range  (s) & Trill Rate Range (Hz) & \# of Birds \\
\hline
Bird A & 400-1200 & 0.6-1.0 & 5-10 & 1 \\
Bird B & 3000-8000 & 0.6-1.0 & 2-6 & 2 \\
Bird C & 2000-10000 & 0.1-0.3 & 4-7 & 1 \\
Bird D & 1000-4000 & 0.2-0.4 &1-3 & 1 \\
Bird E & 3500-7500 & 0.1-0.3& 2-7 & 2 \\
\hline
\end{tabular}
\label{tab:species}
\end{table*}

\subsection{Audio Rendering and Visualization Evaluation}

Visualization and audio analysis are used to examine bird trajectories, spectrograms, and activity timelines, while generated audio files support listening tests and auditory evaluation. These evaluations provide insight into the quality and behavior of the synthesized bird-like sounds in 3D space. 

\subsubsection{Audio Rendering}

 The audio rendering employs digital sound synthesis and multichannel spatialization techniques, ensuring perceptually coherent spatial cues and timbral fidelity during playback. Stereo output is used to preserve the spatialization of bird sounds within the simulated 3D environment. The stereo panning technique is applied to reflect the relative positions of each bird, creating a dynamic representation of their movements across the virtual soundscape. Distance attenuation is also incorporated to simulate changes in volume as birds moved closer to or farther from the listener. This allows for realistic changes in sound intensity and spatial awareness, enhancing the overall auditory experience and ensuring that the 3D spatial cues were faithfully represented in the stereo field.

\begin{figure}[b]
\centering
\includegraphics[width=0.45\textwidth]{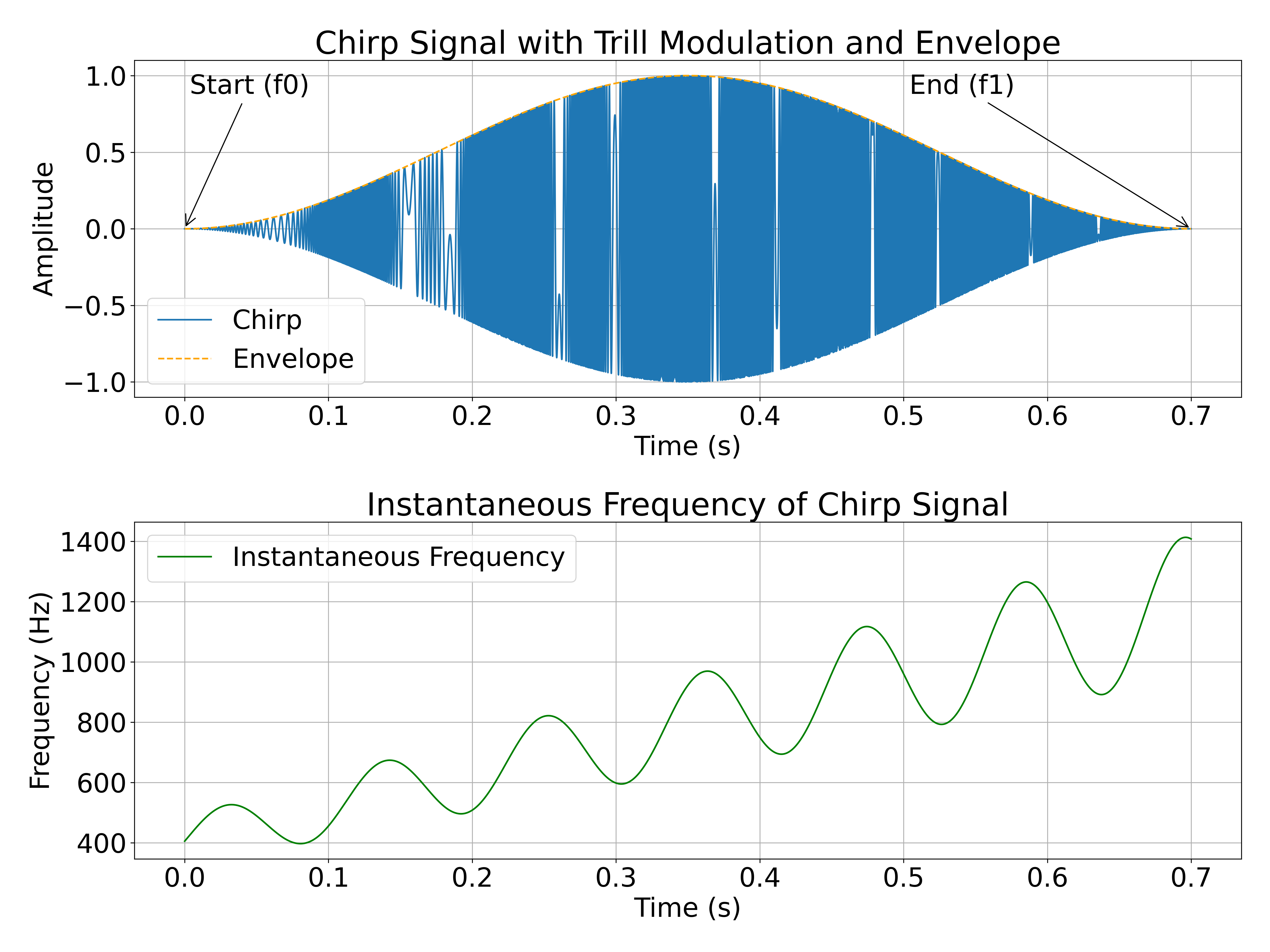}
\caption{Chirp signal and instantaneous frequency derived from acoustic features ($f_0 = 400~\mathrm{Hz}$, $f_1 = 1180~\mathrm{Hz}$, trill rate $= 9~\mathrm{Hz}$, duration $= 0.7~\mathrm{s}$). }
\label{fig:chirp_envelope}
\end{figure}

\begin{figure}[b]
\centering
\includegraphics[width=0.45\textwidth]{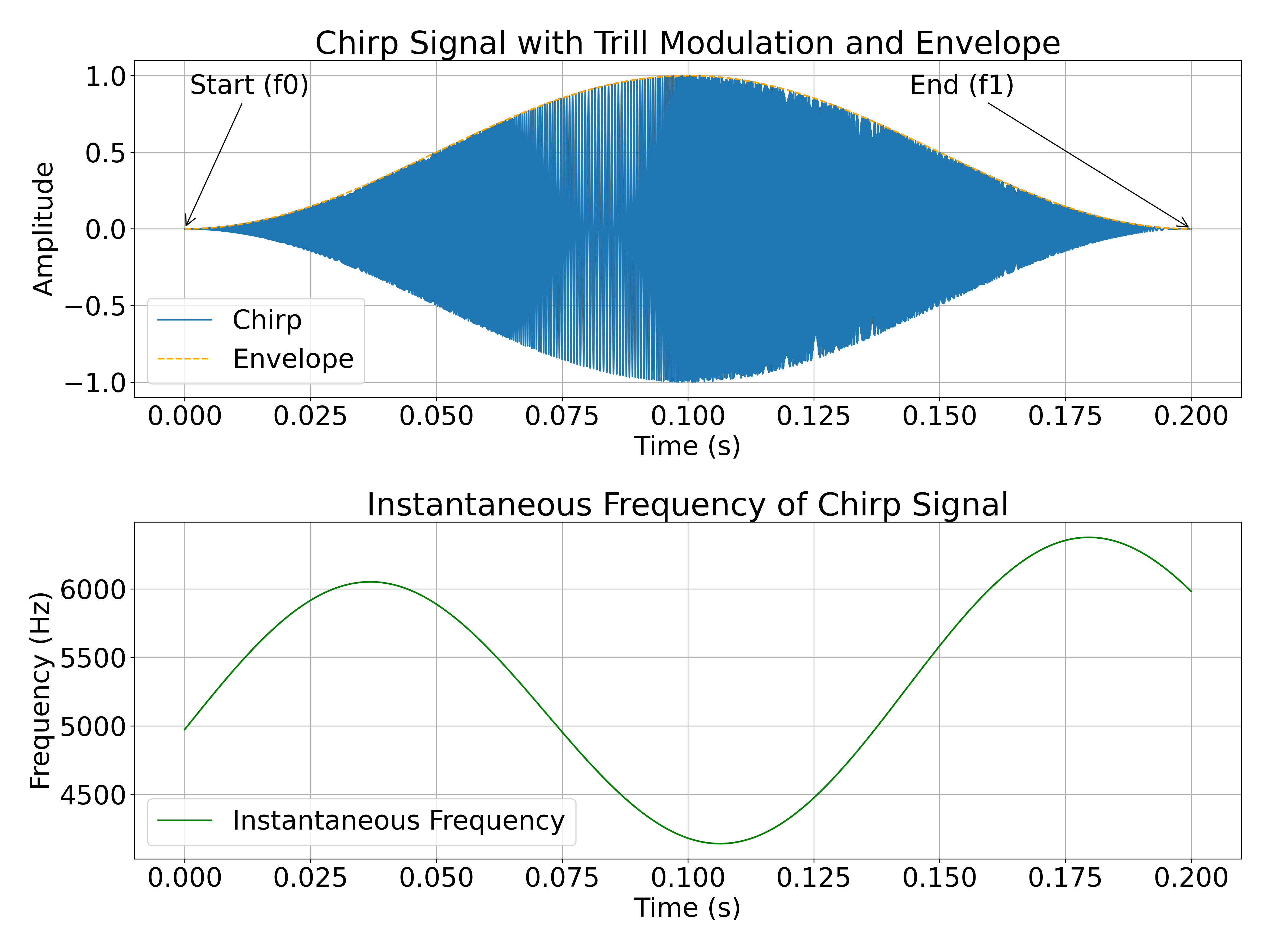}
\caption{Chirp signal and instantaneous frequency derived from acoustic features ($f_0 = 4970~\mathrm{Hz}$, $f_1 = 5350~\mathrm{Hz}$, trill rate $= 7~\mathrm{Hz}$, duration $= 0.2~\mathrm{s}$). }
\label{fig:chirp_envelope_2}
\end{figure}

\begin{figure*}[t]
\centering
\includegraphics[width=\textwidth]{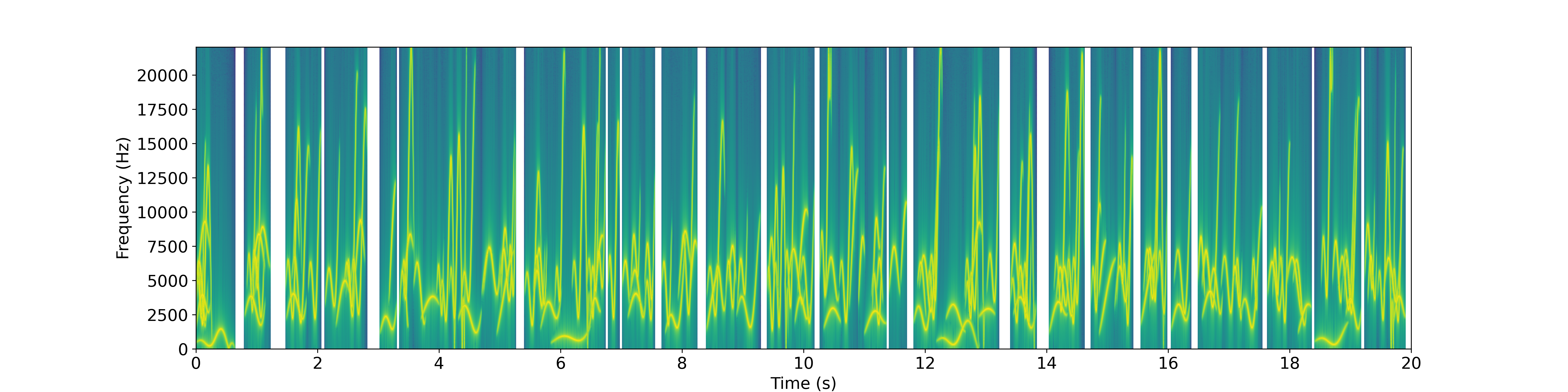}
\caption{Spectrogram of left channel showing overlapping multi-species bird calls.}
\label{fig:spectrogram}
\end{figure*}

\begin{figure*}[t]
\centering
\includegraphics[width=1\textwidth]{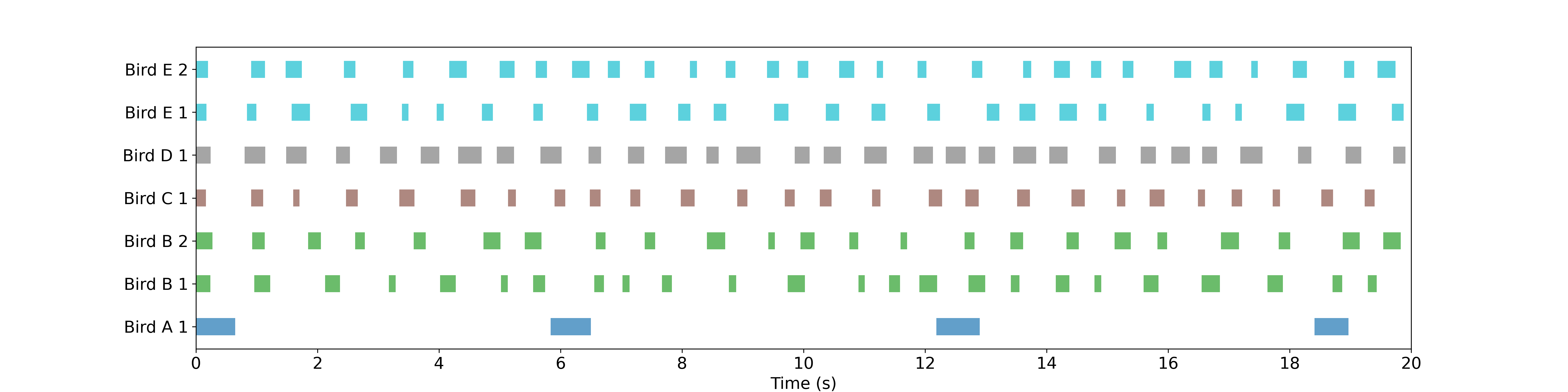}
\caption{Bird chirp activity timelines.}
\label{fig:timeline}
\end{figure*}

\subsubsection{Visualization}

Several complementary visualizations are used to analyze the spatial trajectories and vocal activity of individual birds: 1) \textit{three-dimensional trajectory plots} characterize each bird’s movement within the simulated environment; 2) \textit{spectrograms} provide a detailed view of the temporal and spectral structure of the chirps, including difference between left and right channels, enabling analysis of timbre and spectral evolution; 3) \textit{bird chirp activity timelines} summarize the temporal distribution of vocalizations for each bird over the specified observation period; 4) \textit{time-resolved spectrograms} combines the activity timeline with its corresponding spectrogram, enabling joint inspection of temporal vocal activity and spectral features within a unified multimodal representation; and 5) \textit{comparative analysis of 3D audio waveforms}, specifically the differences between the left and right channels, provides insight into how bird movement within the 3D space influences spatial audio dynamics.  \\

These evaluation techniques provide a comprehensive framework for analyzing spatial dynamics, temporal organization, and species-specific vocal characteristics in the simulated auditory scenes, while grounding the analysis in principles widely used in computer music for rendering and interpreting complex auditory data.

\section{Experiments and Analysis}

The experiments aim to demonstrate the capabilities of the system in generating musically expressive and spatially dynamic bird sounds. Rather than perceptual testing, we emphasize evaluation metrics for bird soundscapes based on both visual and audio representations: 1) \textit{chirp signal and modulation}, showing species-specific trill modulation and amplitude envelopes; 2) \textit{spectrograms of multi-species soundscapes}, illustrating overlapping calls and spectral diversity; 3) \textit{bird chirp activity timelines}, highlighting temporal patterns of vocalizations across species; 4) \textit{per-bird time-resolved spectrograms}, indicating each bird's frequency trajectory and temporal structure, as well as pattern consistency and interactions between individual birds; 5) \textit{3D bird trajectories}, visualizing independent movement patterns of multiple birds within 3D space; 6) \textit{left-right channel spectrogram comparison with differences}, analyzing spatial or source variations in the stereo audio signal; and 7) \textit{comparative analysis of 3D audio waveforms}, accurately capturing bird
movements and positions within the 3D virtual space.

\subsection{Experiment Setup}
The framework was implemented entirely in Python for DSP-based chirp synthesis, 3D spatialization, visualization, and audio rendering. Audio output was generated as WAV files at 44.1 kHz sampling rate for offline playback.

In the experiments, we synthesized 3D soundscapes featuring five species: Bird A, Bird B, Bird C, Bird D, and Bird E. Each species contains one or two birds, each placed at randomized initial positions within the virtual environment.  
Table~\ref{tab:species} summarizes the chirp parameters for each species, including the frequency range, chirp duration range, trill rate range, and the number of birds.


 \def\thefootnote{1}\footnotetext{ \href{https://www.intellisky.org/share/birdsong3D.wav}{Generated birdsong audio demo}.}

The system is capable of generating an extensive variety of birdsong soundscapes, a representative example of which is analyzed in the following subsections. A corresponding WAV audio file$^1$ of the generated soundscape is provided as well.

\subsection{Chirp Signal and Modulation}

Figure~\ref{fig:chirp_envelope} illustrates a synthesized birdsong chirp generated using parameters typical of a short tonal call: starting frequency ($f_{0}$) = 400~Hz, ending frequency ($f_{1}$) = 1180~Hz, trill rate 9~Hz, and duration 0.7 seconds. In the upper panel, the waveform begins with low-frequency oscillations that gradually increase in density as the chirp sweeps upward toward the higher final frequency. A smooth amplitude envelope applies a fade-in at the start and fade-out at the end, producing the symmetric, tapered energy profile characteristic of natural bird vocalizations. Superimposed 9~Hz trill modulation generates periodic clusters of oscillations, creating a rapid vibratory texture. The lower panel shows the instantaneous frequency, revealing the linear sweep from 400~Hz to 1180~Hz along with sinusoidal fluctuations from the trill. These features show how controlled frequency sweeps, amplitude shaping, and structured trill modulation combine to synthesize a naturalistic chirp.

Figure~\ref{fig:chirp_envelope_2} shows another example of generated chirp, using different acoustic features ($f_0 = 4970~\mathrm{Hz}$, $f_1 = 5350~\mathrm{Hz}$, trill rate 7~Hz, duration 0.2~s), along with its corresponding instantaneous frequency. This further demonstrates how the same synthesis principles can generate varied synthetic birdsong elements.

\begin{figure*}[t]
\centering
\includegraphics[width=\textwidth]{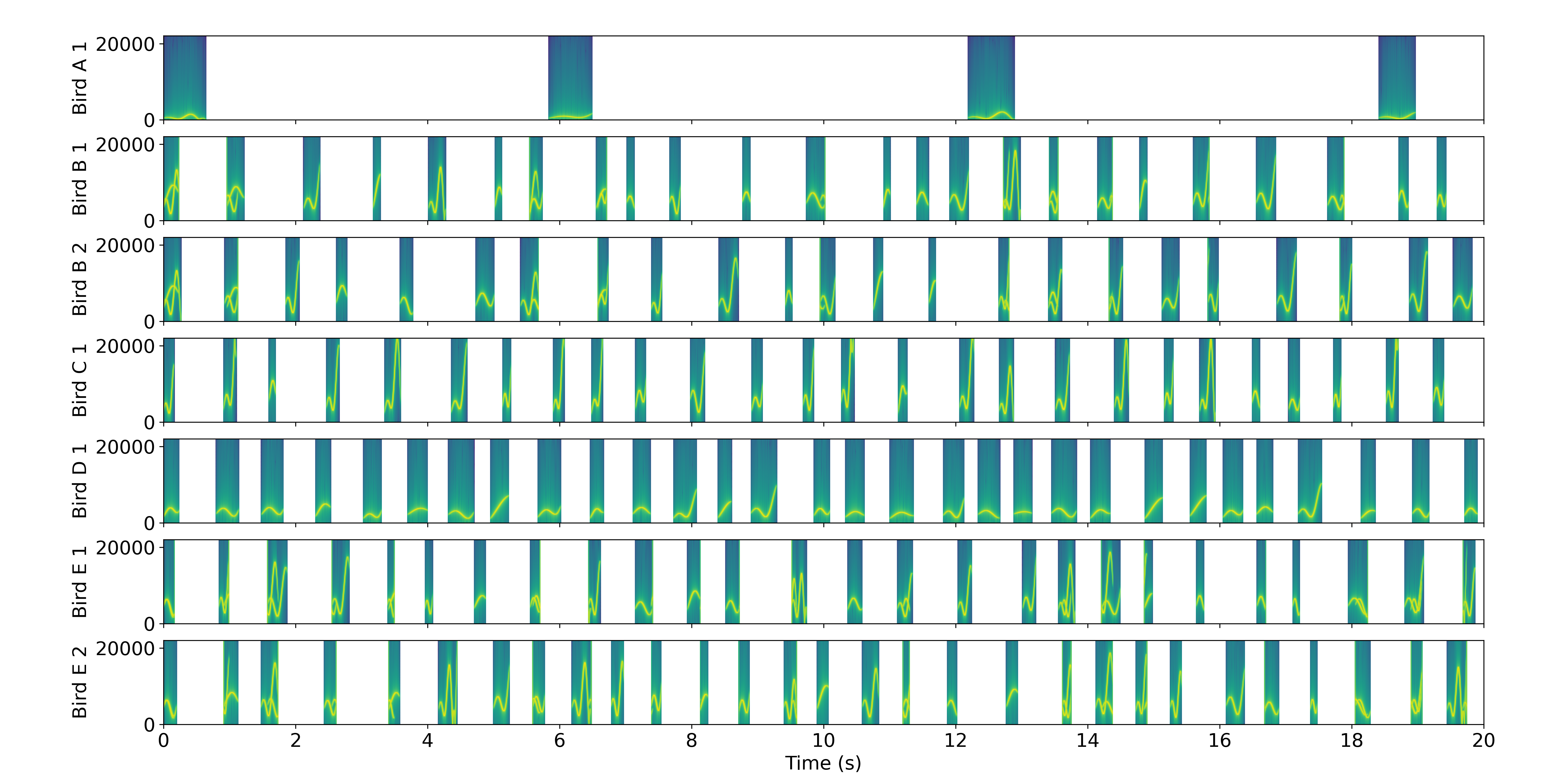}
\caption{Time-resolved spectrograms: chirp activity timeline per bird, demonstrating overlap and temporal patterns.}
\label{fig:timeline_gram}
\end{figure*}

\subsection{Spectrogram of Multi-Species Soundscapes}
The left-channel spectrogram in Figure~\ref{fig:spectrogram} demonstrates the chirp frequencies for each bird of different species. In the figure, the highlighted green lines represent the frequency motion of the 3D soundscape of each of all seven birds, including five bird species. The green lines are measured by frequency, with a high frequency representing high pitches and a low frequency representing a lower pitch. The white spacings of different widths between the blue-green colored spectrogram represent the overall pauses of all the birdsongs sang in the soundscape, mimicking natural birdsong patterns sang by real-world birds. For instance, the first section of the left-channel spectrogram displays two high frequency hill peaks at the beginning and before the pause, as well as a lower frequency wave pattern at the bottom. The diverse changes of the frequency motion of each bird reveal the combination of overlapping bird species songs at a specified time interval in the generated soundscape, revealing a realistic nature soundscape in addition to the 3D trajectories.

\subsection{Bird Chirp Activity Timelines}

Figure~\ref{fig:timeline} displays the bird chirp activity timeline for all seven birds across five species. Each species is assigned a distinct color, such as cyan for the Bird E species. The length of each rectangle represents the duration of an individual bird chirp's, mapped to the specified time scale on the x-axis. The gaps between the rectangles indicate the duration of pauses between chirps, mimicking the natural patterns observed in real-world bird vocalizations. Birds of the same species tend to exhibit similar chirp lengths and pause intervals; for example, Bird E 1 and 2 show comparable chirp patterns, while Bird A 1 and Bird C 1 display notably contrasting chirp characteristics. The variation in both chirp and pause duration not only reflects the natural dynamics of bird vocalization patterns, but also captures species-specific differences in frequency range and chirp duration.

\subsection{Time-Resolved Spectrograms}

Figure~\ref{fig:timeline_gram} presents detailed per-bird chirp timelines alongside time-resolved spectrograms, providing a comprehensive visualization of vocal activity across individuals. These representations not only capture the precise timing and duration of each bird’s chirps, but also reveal temporal overlaps and potential interactions between birds. By aligning the chirp events with their corresponding spectral features, the figure highlights patterns of coordination, call overlap, and sequential calling, offering insights into the dynamics of acoustic communication within the group.

Moreover, the spectrograms for each bird indicate that the frequency trajectories form smooth or rapid pitch sweeps, closely resembling the expressive contours typical of stylized bird calls. The temporal organization is also well articulated, with distinct onsets, clean offsets, and natural inter-event spacing that mirror the rhythmic structure found in real bird-like vocal gestures. Examining multiple generated examples shows strong pattern consistency, indicating that the synthesis method reliably produces stable and coherent chirp structures rather than random or noisy artifacts. Furthermore, no discontinuities, energy smearing, or unexpected spectral bands are observed, indicating that the synthesis system generates robust, artifact-free outputs. Overall, the visual analysis demonstrates that the generated sounds are convincingly bird-like, with preserved clarity, structural coherence, and reproducibility.

\subsection{3D Bird Trajectories}

\begin{figure}[h]
\centering
\includegraphics[width=0.40\textwidth]{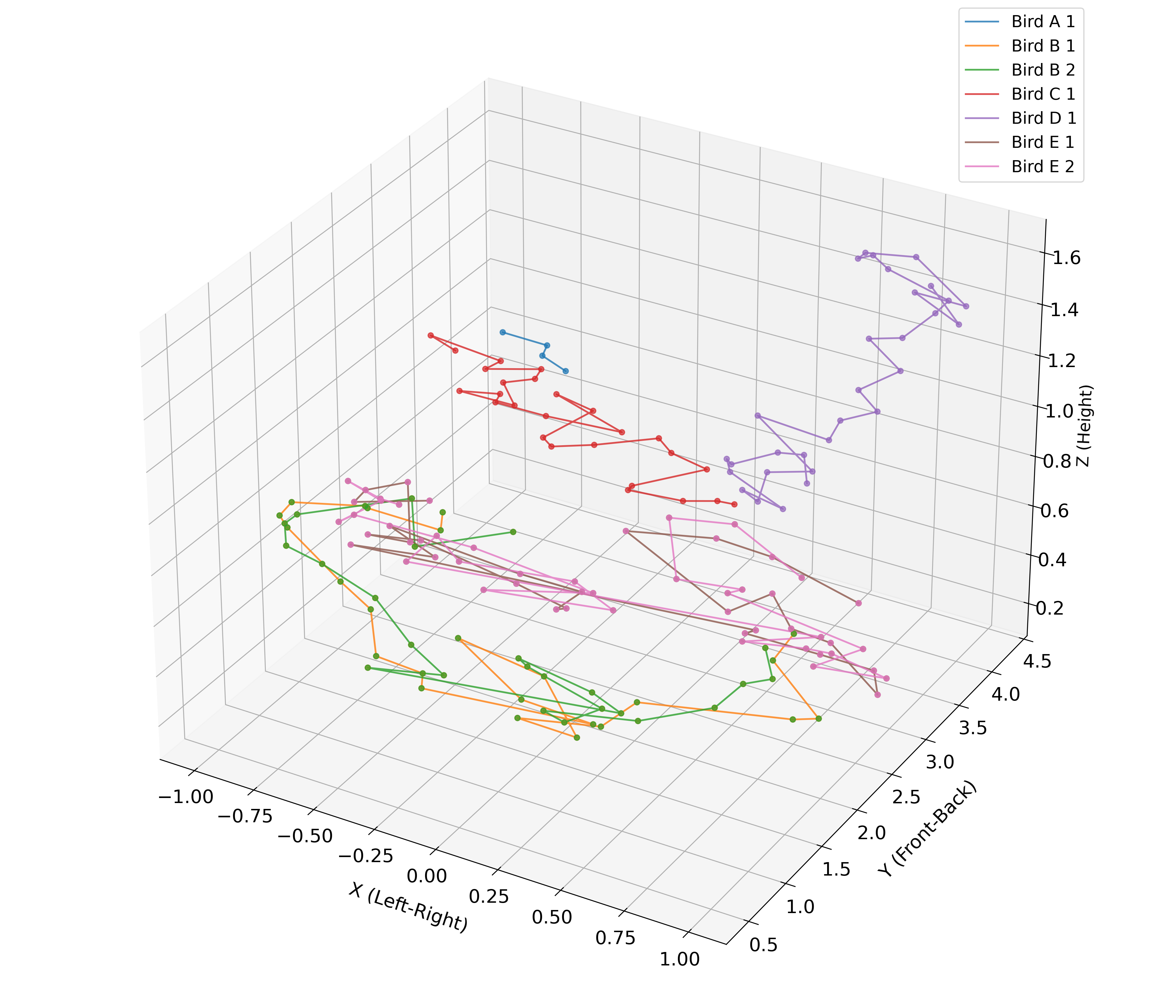}
\caption{3D trajectories of each bird.}
\label{fig:trajectory}
\end{figure}

\begin{figure*} [t]
\centering
\includegraphics[width=\textwidth]{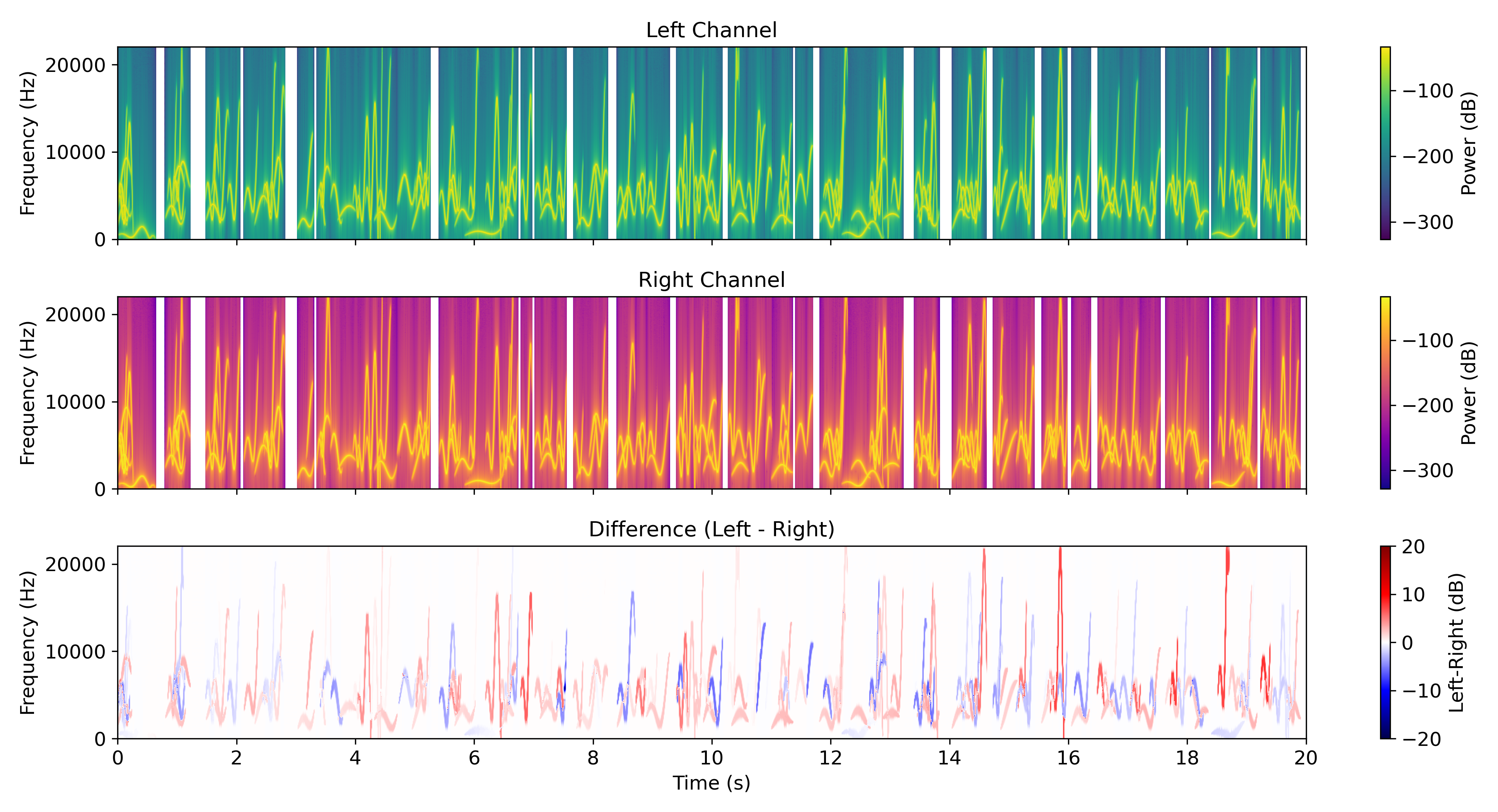}
\caption{Comparative analysis of left and right channel spectrograms with highlighted differences.}
\label{fig:gram_diff}
\end{figure*}

\begin{figure*} [t]
\centering
\includegraphics[width=0.98\textwidth]{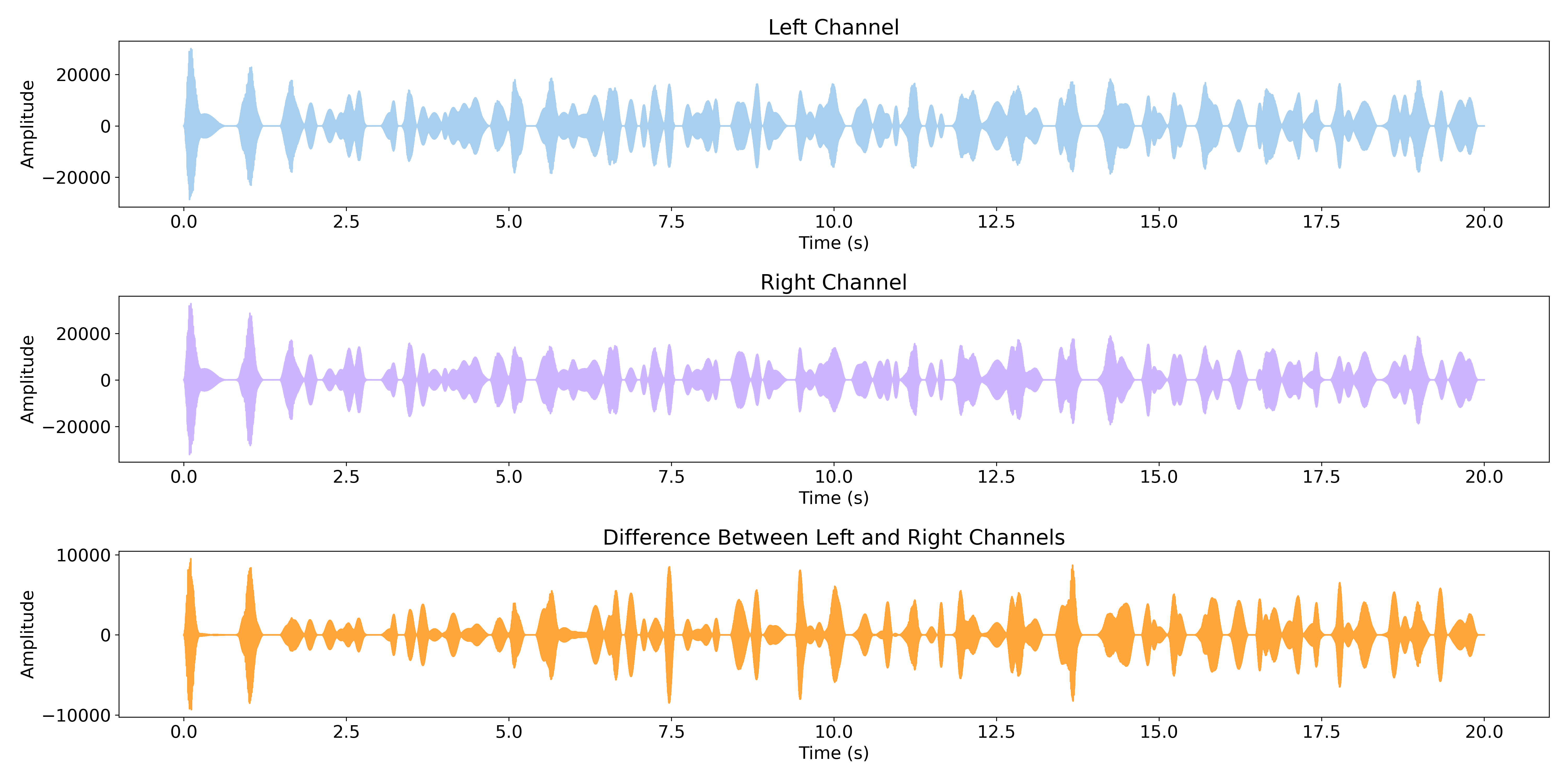}
\caption{Comparative analysis of left and right stereo channels with highlighted differences.}
\label{fig:wav_diff}
\end{figure*}

Figure~\ref{fig:trajectory} illustrates 3D bird movements over time. Each color line represents each bird, as indicated in the key. The 3D scale shows the spatial location of each bird, contributing to the 3D stereo effect, with the audio output originating from the center of the 3D space. For example, the path of Bird E 1, ranging from -0.1-0.8 on the x-axis and 0.3-1.5 in height, indicates that its birdsong will be heard from the  middle right in the 3D stereo output. In contrast, the path of Bird B 1, ranging from -1.0 to 0.75 on the x-axis, results in its birdsongs moving across both the left and right channels, as the bird flies from left to right (or vice versa), with the sound positioned in the middle of the 3D stereo field.

This figure demonstrates the spatialization capabilities of the framework, showing how individual bird trajectories are reconstructed within the simulated auditory scene. These trajectories serve as spatial control parameters for dynamic source positioning. The figure also highlights the support of the system for complex, non-linear movement patterns, with each bird following a unique path on the xy-plane and varying elevation along the z-axis. This variability is essential for maintaining spatial clarity, even in overlapping motion scenarios. The examples show that the framework generates rich positional data necessary for realistic 3D auditory scene synthesis, aligning with spatialization techniques commonly used in computer music and sound design.

\subsection{Left-Right Channel Spectrogram Comparison}
In Figure~\ref{fig:gram_diff}, the three-panel spectrogram visualization effectively compares the left and right audio channels along with their difference over a 20-second interval. The top panel displays the frequency content of the left channel, showing prominent activity concentrated primarily below 7,500 Hz, with occasional higher-frequency bursts reaching up to around 20,000 Hz. The middle panel shows the right channel with a similar pattern and frequency distribution but subtle variations in intensity and timing, as indicated by slightly different spectral shapes and brightness. The intensity colormaps of both left and right channels show dynamic variations, with energy levels ranging from approximately $-320$~dB to $-100$~dB. The bottom panel depicts the differential spectrogram, highlighting the discrepancies between channels, using a diverging red-blue colormap to indicate positive and negative intensity differences up to $\pm 20$~dB. This difference plot effectively emphasizes localized spectral disparities, suggesting channel-specific variations potentially arising from stereo recording characteristics or spatial filtering effects.



\subsection{3D Audio Rendering}
To evaluate the effectiveness of the audio rendering approach, we synthesized multiple 20-second soundscapes featuring five species. The birds’ vocalizations were generated based on the species-specific parameters, ensuring that the acoustic patterns, temporal modulation, and spatial movement were accurately represented. In the experiments, the resulting audio was played back in stereo, with panning and distance attenuation applied in real time based on the birds’ 3D trajectories. The positioning and movement of each bird within the simulated space were also visualized to assess the spatial coherence of the soundscapes as shown in Figure \ref{fig:trajectory}. The experiments focused on ensuring that the synthesized bird sounds preserved perceptual accuracy in both auditory spatialization and naturalistic variation of  vocal patterns.

\subsection{Comparative Analysis of 3D Audio Waveforms }
Figure~\ref{fig:wav_diff} presents a comparative visualization of the stereo audio signal by plotting the left channel, right channel, and their sample-wise difference over a 20-second duration  for a multi-bird soundscape consisting of several species. The left and right channels display similar structures with subtle variations in amplitude and timing, reflecting the movement of multiple birds within the 3D space. These differences, highlighted in the difference plot (orange), correspond to spatial effects such as panning and phase shifts caused by birds moving at different positions and distances relative to the listener. For example, when birds move from left to right, the right channel’s amplitude increases while the left channel’s decreases. The difference plot reveals how overlapping bird vocalizations are distributed across the stereo field, confirming the ability of the system to accurately render dynamic, multi-bird multi-species soundscapes with spatial depth.

This comparison demonstrates that the synthesized bird soundscapes maintain a coherent spatial structure, with the left-right channel variations accurately reflecting the birds' movements, positions, and distance relationships within the 3D virtual environment.

\section{Conclusion and Future Work}

We present an innovative, fully algorithmic DSP-based framework for generating dynamic 3D multi-species bird soundscapes that incorporate species-specific acoustic patterns, temporal chirp modulation, overlapping choruses, inter-bird variability, and 3D spatialization. The system produces clear, artifact-free birdsongs with independently moving birds per species, fully controllable chirp sequences, and realistic spatial effects, achieving greater flexibility and fidelity than existing algorithmic, data-driven, or hand-designed techniques. The system also includes visualization and analytical tools, such as bird trajectories, chirp activity timelines, spectrograms, and multi-channel audio waveforms, that support creative exploration and in-depth analysis of spatial and temporal dynamics. Visual and audio evaluations demonstrate the potential of the framework for immersive ecological simulations, virtual reality applications, and bioacoustic research.

Future work will focus on enhancing environmental realism by incorporating factors such as weather effects and integrating real-time interactivity for dynamic sound generation. Additionally, we aim to expand the application of the system in generative composition and ecoacoustic music, enabling composers to create context-aware, interactive soundscapes based on species interactions and spatial dynamics.

\section{Ethical Implications}

Multi-species bird soundscape synthesis often raises ethical issues, such as arising from the use of recordings or training data. Our approach, however, generates dynamic bird soundscapes with purely algorithmic methods without reliance on recordings, preventing any ethical and copyright concerns in the creation of 3D environment soundscapes. 

\bibliographystyle{IEEEtran}
\bibliography{references}

\end{document}